# What is transiting HD 139139 ?
*Jean Schneider*
*Paris Observatory*

The NASA mission Kepler has detected 28 transits with depths $\eta \approx 220$ ppm and durations $D_T \approx 2.5$ hours in the light curve of HD139139 (radius $R_* = 1.14\, R_\odot$, $M_* = 1\, M_\odot$ and $d_* = 100$ pc) during a 87 days campaign . Their arrival times are erratic. Rappaport et al. (2019) discard ten explanations. It is not clear if the transits are for HD 139139 or for a star B at 3.3 ". New radial velocity variation data give RV <10 m/s for HD139139 in 4 days (F. Bouchy and S. Udry, private communication) excluding a close-in $M = 50\, M_{Jup}$ planet orbiting HD 139139, as proposed earlier (Schneider 2019). (However, this explanation is still valid for the star B for which radial velocity data are very poor). Here I thus explore new tentative explanations and their likelihood: 1/ An eccentric transiters belt around HD 139139 2/ Interstellar transiters 3/ Solar System objects.

I consider cases where there are several objects transiting the star HD 139139. Details are in preparation. As a general constraint, any transiter must have an orbital period larger than 87/2 = 43 days to avoid 3 transits with equal interarrival times in 87 days. Since a transiter can make only two transits in 87 days, there are 28/2 = 14 transiters.

## *1 Eccentric transiters belt*

Objects transiting HD139139 at the periastron of an orbit with an orbital period $P_T$ and an eccentricity *e* give a mean transit duration

$$D_T = \sqrt{3}\, R_* \sqrt{(1-e)/(1+e)} \left(2\pi\, P_T / GM_*\right)^{1/3} \quad (1)$$

Several objects randomly distributed on an asteroid-like belt give random transit arrival times. From equation (1)

$$e = \left(1 - D_T^2\, GM_* / a_T\, R_*^2\right) / \left(1 + D_T^2\, GM_* / a_T\, R_*^2\right) \quad (2)$$

To have $D_T = 2.5$ hours, *e* must be 0.74 for a 43 days orbit. For a belt of trojan objects with an azimuthal distribution of 40° similar to Solar System trojans (Figure 1a), the orbital period must be 87 days × 40°/360° = 783 days, leading to an eccentricity of 0.95 (for comparison, $e = 0.97$ for HD 20782b). According to Lyra et al. (2009), trojan objects can have several Earth masses.

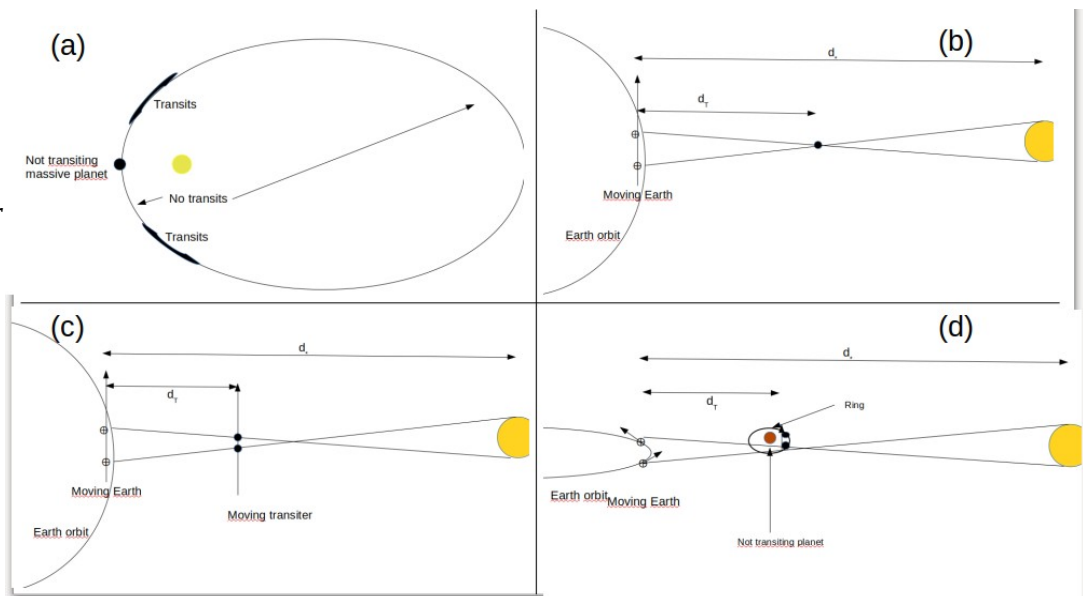

*Figure 1* Four configurations of transiters

The transits should then disappear when the trojan belt is not transiting and reappear after a time $P_T(80°/360°)$ or $P_T(200°/360°)$. However, one then faces stability problems making this configuration less likely.

## 2 Interstellar objects

Transiters at rest at a distance $d_T \ll d_*$ give a transit duration, due to the Earth velocity $V_\oplus$ around the Sun, of $D_T = d_T \times 2 R_*/V_\oplus d_*$ or $d_T = D_T V_\oplus d_*/2 R_*$ (Figure 1b). For a 2.5 hour transit $d_T = 20$ pc. Then the 220 ppm depth gives a radius of $\sqrt{\eta} R_* d_T/d_* = 0.3 R_\oplus$, or a mass of $0.02 M_\oplus$. The group of objects can either be a series of objects orbiting a non-transiting faint star or brown dwarf, or a self-gravitating spherical cluster similar to stellar globular clusters. However, in the latter case one would have to explain its formation. If the group of objects has a typical velocity of 20 km/s, the numbers remain the same within a factor 2.

Suppose that, according to the virial theorem to maintain its stability, the cluster of transiters has the same velocity dispersion $\sqrt{GM_{cl}/r_c} \approx 2$ km/s (where $M_{cl}$ is its total mass), as a typical stellar globular cluster (Meylan & Heggie 1997).

Then with a core radius of $r_c = 0.5 \times 87\,\text{days}/V_\oplus = 1.5$ AU, the total mass of the cluster of transiters should be $10^{-2} M_\odot$. The mass of each transiter being $\sim 2\,10^{-2} M_\oplus$, the cluster should contain $\sim 3\,10^5$ objects.

For a core radius of $r_c$ of 1.5 AU, the projected interdistance of objects is $r_c/(2\,10^8)^{1/3} \approx 5 R_\odot$. The interarrival time of transits then is $5 R_\odot/V_\oplus = 2.5\,10^5$ sec, in agreement with the observed mean interarrival times of 3 days.

## 3 Solar System objects

Let us take objects at a distance $d_T = 500$ AU or more. Then their size $r_T$ is $\sqrt{\eta} R_* d_T/d_* = 17$ m. Their orbital velocity around the Sun, at more than 500 AU, is less than 1.3 km/s.

The duration $D_T$ of their transits is thus dominated by the Earth velocity on its orbit around the Sun: $D_T = 2 R_* (d_T/d_*)/V_\oplus$. For $d_T = 500$ AU, the duration is 12 sec, incompatible with the observed 2.5 h duration.

But one can assume that there is some source of extra high velocity $V_T$ of each of these objects which compensate the Earth velocity (Figure 1c).

After some algebra one finds that (since $d_T \ll d_*$) the transverse velocity of transisters is
$$V_T = V_\oplus - 2 d_T R_*/(d_* \times 2\,\text{hours}) \sim 30 - \epsilon\,\text{km/s}$$
(where $\epsilon = 2 d_T R_*/(D_T V_\oplus d_*)$ is negligible compared to 30 km/s).

Suppose a configuration where there is a group of transiters. Then, since there are transits during 87 days at least, this group, supposed to have a group orbital velocity $V_G$ around the Sun of 1.3 km/s at 500 AU, must have a transverse extension $a_T = 87\,\text{days} \times V_G = 0.065$ AU at least.

As a concrete model, these objects could be in a ring of ~ 17 m rocks around a massive, yet unseen, planet at > 500 AU (Figure 1d). For circular orbits around a planet, to have a velocity of ~ 30 km/s for the transiters [1] at $a_T/2 = 0.03$ AU around their parent planet, the latter must have a mass of 30 Jupiter masses (Figure 1d).

Or the transiters could presently be at the perihelion of an orbit with an eccentricity $e$. To have a velocity of 30 km/s with a semi-major axis of 500 AU, from equation (2), $e$ must be 0.9998, similar to the orbit of the comet C/1680 V1 (semi-major axis 444 AU, eccentricity 0.999996). Since one must have at least 28/2 = 14 transiters, they could be the result of a disrupted comet or asteroid by collision with an interstellar asteroid entering the Solar System (Couture 2019, Moro-Martin et al. 2009).

---

1. To be more precise, the velocity then is $30\,km/s \times \cos\omega$ where $\omega$ is a random orbital phase factor around the planet; it leads to the observed dispersion of transit durations. I skip this discussion here.

Finally, we clearly need more data (more transits, radial velocities and imaging).

**Acknowledgement**
I thank Valéry Lainey for discussions.